\documentclass{PoS}

\title{Lattice QCD Analysis for Gluons}

\ShortTitle{Lattice QCD Analysis for Gluons}

\author{\speaker{Hideo Suganuma}, Takumi Iritani, Arata Yamamoto \\
        Department of Physics, Kyoto University, 
Kitashirakawaoiwake, Sakyo, Kyoto 606-8502, Japan\\
        E-mail: \email{suganuma@scphys.kyoto-u.ac.jp}
}

\author{Hideaki Iida \\
The Institute of Physical and Chemical Research (RIKEN), 
Wako, Saitama 351-0198, Japan}

\abstract{
Nonperturbative properties of gluons are studied in SU(3) lattice QCD 
at the quenched level. 
The first subject is a functional-form analysis of 
the gluon propagator $D_{\mu\nu}^{ab}(x)$ in the Landau gauge. 
We find that the gluon propagator 
$D_{\mu\mu}^{aa}(r)$ obtained in lattice QCD 
is well described by the four-dimensional (4D) Yukawa-type function 
$e^{-mr}/r$ with $m \simeq 600$MeV 
for the Euclidean 4D distance $r = 0.1 \sim 1.0$ fm. 
In momentum space, the gluon propagator 
$\tilde D_{\mu\mu}^{aa}(p^2)$ ($p= 0.5 \sim 3$ GeV) is found to be 
well approximated with a new-type propagator of $(p^2+m^2)^{-3/2}$, 
which corresponds to 4D Fourier image of 
the Yukawa-type function. 
Associated with the Yukawa-type gluon propagator, 
we derive analytical expressions for 
the zero-spatial-momentum propagator $D_0(t)$, 
the effective mass $M_{\rm eff}(t)$, and the spectral function 
$\rho(\omega)$ of the gluon field. 
The mass parameter $m$ turns out to be the infrared effective mass of gluons. 
The obtained gluon spectral function $\rho(\omega)$ 
is almost negative-definite for $\omega >m$, 
except for a positive $\delta$-functional peak at $\omega=m$.
The second subject is a lattice-QCD determination of 
the relevant gluonic momentum-component for color confinement. 
As a remarkable fact, the string tension is found to be almost unchanged 
even after cutting off the high-momentum gluon component above 1.5 GeV 
in the Landau gauge. 
In fact, the relevant gluonic scale for color confinement 
is concluded to be below 1.5 GeV. 
}

\FullConference{International Workshop on QCD Green's Functions, 
Confinement and Phenomenology\\
                 September 7-11 , 2009\\
                 ECT Trento, Italy}

\begin{document}

\section{Introduction}

Quantum chromodynamics (QCD) and the gluon field 
$A_\mu(x) =A_\mu^a(x)T^a \in {\rm su(3)}$ 
were first proposed by Nambu in 1966 \cite{N66} 
just after the introduction of color degrees of freedom. 
Although QCD has been established as the fundamental gauge theory 
of the strong interaction with many successes, 
there are still unsolved problems 
on nonperturbative QCD in the low-energy region.

The analysis of gluon properties is an important key point 
to clarify the nonperturbative aspects of QCD \cite{C8207}.
In particular, the gluon propagator, {\it i.e.}, 
the two-point Green function is one of the most basic quantities in QCD, 
and has been investigated with much interests in various gauges, 
such as the Landau gauge 
\cite{MO87,GGKPSW87,BPS94,MMST9395,C9798,UK,
AU,LRG02,BIMPS090705,SO0607,ABP08,ISI09}, 
the Coulomb gauge \cite{GOZ0304,CZ02}, 
and the maximally Abelian (MA) gauge \cite{AS99,K9800}, 
in the context of various aspects of QCD.
Dynamical gluon-mass generation \cite{C8207,B8283} 
is also an important subject 
related to the infrared gluon propagation. 
While gluons are perturbatively massless, 
they are conjectured to acquire a large effective mass 
as the self-energy through their self-interaction 
in a nonperturbative manner.
For example, 
glueballs, color-singlet bound states of gluons, 
are considered to be fairly massive, 
{\it e.g.}, about 1.5GeV for the lowest $0^{++}$ and about 2GeV 
for the lowest $2^{++}$, 
as indicated in lattice QCD calculations \cite{R05,ISM02}. 

In this paper, we study the functional form of the gluon propagator 
in the Landau gauge in SU(3) lattice QCD Monte Carlo calculations, 
especially for the infrared and intermediate region of $r=0.1 \sim 1.0$fm, 
which is relevant for the quark-hadron physics \cite{R05,IOS05}, 
and aim at a nonperturbative description of gluon properties, 
based on the obtained function form of the gluon propagator.
As another subject, using lattice QCD, we also study 
the relevant gluonic momentum-component for color confinement 
at the quantitative level, 
by introducing a cut in the momentum space 
\cite{YS0809}.

\section{Formalism for gluon propagator in Landau gauge}
In this section, we briefly review the formalism of Landau gauge fixing
and the gluon propagator in Euclidean space-time.
The Landau gauge is one of the most popular gauges in QCD, 
and keeps Lorentz covariance and global ${\rm SU}(N_c)$ symmetry.
Owing to these symmetries and the transverse property, 
the color and Lorentz structure of the gluon propagator 
is uniquely determined.

In Euclidean QCD, the Landau gauge has a global definition 
to minimize the global quantity, 
\begin{equation}
R \equiv \int d^4 x \ {\rm Tr} \{A_\mu(x) A_\mu(x)\} 
= \frac{1}{2} \int d^4 x A_\mu^a(x) A_\mu^a(x),
\end{equation}
by gauge transformation.
The local condition $\partial_\mu A_\mu(x) = 0$ 
is derived from the minimization of $R$.
The global quantity $R$ can be regarded as 
``total amount of the gauge-field fluctuation'' in Euclidean space-time. 
In the global definition, 
the Landau gauge has a clear physical interpretation 
that it maximally suppresses artificial gauge-field fluctuations 
relating to gauge degrees of freedom \cite{ISI09}.
We then expect that only the minimal quantum fluctuation of 
the gluon field survives in the Landau gauge, 
and the physical essence of gluon properties can be investigated 
without suffering from artificial fluctuations of 
gauge degrees of freedom.

In lattice QCD, the gauge field is described by 
the link-variable $U_\mu(x) \equiv e^{iagA_\mu(x)}$ with 
the lattice spacing $a$ and QCD gauge coupling $g$. 
The Landau gauge is defined by the maximization of 
\begin{equation}
R_{\rm latt} \equiv \sum_x \sum_\mu {\rm Re} {\rm Tr} U_\mu(x),
\end{equation}
by the gauge transformation. 
The maximization of $R_{\rm latt}$ 
corresponds to the minimization of $R$ in the continuum theory, 
and maximally suppresses the gauge-field fluctuation.

The bare gluon field $A^{\rm bare}_\mu(x) \in {\rm su}(N_c)$ 
is defined from the link-variable as 
\begin{equation}
A^{\rm bare}_\mu(x) \equiv \frac{1}{2iag} 
\left[ U_\mu(x) - U_\mu^\dagger(x) \right] 
- \frac{1}{2iagN_c} {\rm Tr} \left[ U_\mu(x) - U_\mu^\dagger(x) \right],
\label{eq:gluonfield}
\end{equation}
where the second term is added to make $A_\mu^{\rm bare}$ traceless.
In the Landau gauge, the minimization of gluon-field fluctuations 
justifies the expansion by small lattice spacing $a$. 
The renormalized gluon field $A_\mu(x)$ is obtained by multiplying 
a real renormalization factor $Z_3^{-1/2}$ as 
$
A_\mu(x) \equiv Z_3^{-1/2} A^{\rm bare}_\mu(x).
$

In the Euclidean metric, 
the gluon propagator $D_{\mu\nu}^{ab}(x)$ is defined 
by the two-point function as  
\begin{equation}
D_{\mu\nu}^{ab}(x,y) \equiv \langle A_\mu^a(x) A_\nu^b(y) \rangle
=D_{\mu\nu}^{ab}(x-y).
\end{equation}
In the coordinate space, we investigate 
the scalar combination of the gluon propagator 
\begin{equation}
D(r) \equiv \frac{1}{3 (N_c^2-1) } D_{\mu\mu}^{aa}(x)
     = \frac{1}{3 (N_c^2-1) } 
\langle A_\mu^a(x)A_\mu^a(0)\rangle,
\label{eq:defPropagator}
\end{equation}
as a function of the 4D Euclidean distance 
$r \equiv |x| \equiv (x_\mu x_\mu)^{1/2}$.

\section{Lattice QCD result for the functional form of 
gluon propagator in Landau gauge}

We study the functional form of 
the gluon propagator $D(r)\equiv D^{aa}_{\mu\mu}(r)/24$ 
in the Landau gauge in SU(3) lattice QCD, 
in the infrared and intermediate region of 
$r \equiv (x_\alpha x_\alpha)^{1/2} = 0.1 \sim 1.0$fm, 
which is the relevant scale of quark-hadron physics.
Here, we mainly deal with the coordinate-space 
gluon propagator $D(r)$, which is directly obtained 
from lattice calculations and then more primary 
than the momentum-space propagator 
$
\tilde D(p^2)= \int d^4x \ e^{ip \cdot x} D(x).
$
The ${\rm SU}(3)$ lattice QCD Monte Carlo calculations 
are performed at the quenched level 
using the standard plaquette action with 
$\beta \equiv 2N_c/g^2$=5.7, 5.8, and 6.0, 
on the lattice size of 
$16^3 \times 32$, $20^3 \times 32$, and $32^4$, respectively.
The lattice spacing $a$ is found to be $a = 0.186, 0.152$, and $0.104$fm, 
at $\beta$ = 5.7, 5.8, and 6.0, respectively, 
when the scale is determined so as to reproduce the string tension as 
$\sqrt{\sigma} = 427$MeV 
from the static Q$\bar {\rm Q}$ potential \cite{STI04}.
Here, we choose the renormalization scale at $\mu=4{\rm GeV}$ for $\beta=6.0$ 
\cite{UK,AU}, and make corresponding rescaling for $\beta$=5.7 and 5.8.

Figure 1(a) and (b) show the coordinate-space gluon propagator $D(r)$ and 
the momentum-space gluon propagator $\tilde D(p^2)$, respectively.
Our lattice QCD result of $\tilde D(p^2)$ is 
consistent with that obtained in the previous lattice studies \cite{UK,AU},
although recent huge-volume lattice studies \cite{BIMPS090705,SO0607} 
indicate a suppression of the gluon propagator 
in the Deep-IR region ($p < 0.5$GeV),
compared with the smaller lattice result.

\begin{figure}[h]
\vspace{-0.05cm}
\begin{center}
\includegraphics[scale=0.9]{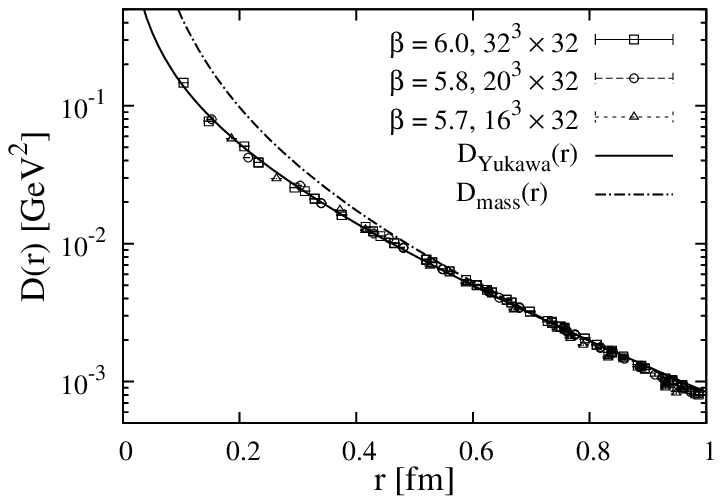}
\includegraphics[scale=0.9]{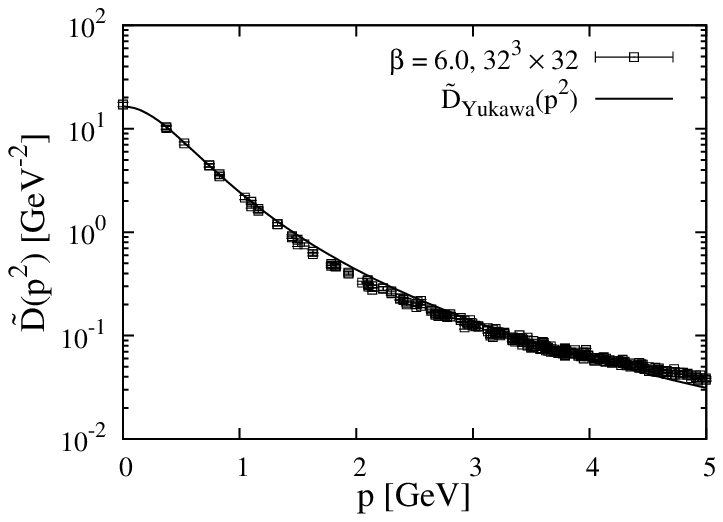}
\caption{
(a) Lattice QCD data of the Landau-gauge gluon propagator 
$D(r)\equiv D_{\mu\mu}^{aa}(x)/24$ 
in coordinate space 
at $\beta$=5.7, 5.8, and 6.0, 
and the Yukawa-type function 
$D_{\rm Yukawa}(r)=Ame^{-mr}/r$~ (solid line) 
with $m=0.624$GeV and $A=0.162$.
The dash-dotted line denotes 
a typical example of the massive-vector propagator $D_{\rm mass}(r)$.
(b) The Landau-gauge gluon propagator 
$\tilde{D}(p^2)=\tilde{D}_{\mu\mu}^{aa}(p^2)/24$ 
in momentum space. 
The symbols denote the lattice QCD data at $\beta = 6.0$, 
where the momentum is defined as 
$p_\mu=\frac{2}{a}\sin(\frac{\pi n_\mu}{L_\mu})$.
The solid line denotes 
the Yukawa-type propagator in the momentum space, {\it i.e.}, 
$\tilde{D}_{\rm Yukawa}(p^2) = 4\pi^2 Am (p^2+m^2)^{-3/2}$.
}
\end{center}
\vspace{-0.05cm}
\end{figure}

First, we consider 
the coordinate-space propagator of the free massive-vector field \cite{AS99}, 
\begin{eqnarray}
D_{\rm mass}(r) = \int \frac{d^4p}{(2\pi)^4} e^{-ip\cdot x}\frac{1}{p^2+m^2} 
= \frac{1}{4\pi^2} \frac{m}{r} K_1(mr),
\label{eq:Dmassive}
\end{eqnarray}
with the modified Bessel function $K_1(mr)$. 
For large $r$, one finds 
$
K_1(mr) \simeq \sqrt{\frac{\pi}{2mr}} e^{-mr},
$
and the massive propagator behaves as
$D_{\rm mass}(r) \sim r^{-3/2}e^{-mr}$.
In Fig.1(a), we add the fit result of the lattice data with $D_{\rm mass}(r)$ 
in the fit-range of $r = 0.6 \sim 1.0$fm.
In the IR region, this fit seems well, 
and the effective mass $m$ is estimated to be about 500MeV. 
However, as shown in Fig.1(a), the lattice gluon propagator $D(r)$ 
cannot be described with $D_{\rm mass}(r)$ 
in the whole region of $r = 0.1 \sim 1.0$fm.

By the functional-form analysis of the gluon propagator, 
we find that 
the Landau-gauge gluon propagator $D(r)$ in the coordinate space 
is well described by the 4D Yukawa-type function \cite{ISI09}
\begin{equation}
D_{\rm Yukawa}(r)=Am\frac{e^{-mr}}{r}
\end{equation} 
with $m=0.624(8)$GeV and $A=0.162(2)$ 
in the range of $r=0.1 \sim 1.0$fm, 
as shown in Fig.1(a).

In Fig.1(b), we add by the solid line the Fourier transformation of 
the Yukawa-type function $D_{\rm Yukawa}(r)$, {\it i.e.},  
$\tilde{D}_{\rm Yukawa}(p^2) = 4\pi^2 Am (p^2+m^2)^{-3/2}$,
with the same parameters $m = 0.624$GeV and $A= 0.162$ as 
those used for the coordinate-space gluon propagator.
The lattice QCD data of $\tilde{D}(p^2)$ 
are found to be approximated with $\tilde{D}_{\rm Yukawa}(p^2)$ 
in the range of $p \le 3{\rm GeV}$, while 
they seem to be consistent with the tree-level massless propagator 
$\tilde D_{\rm tree}(p^2) =1/p^2$ for large $p^2$.

We summarize the functional form of 
the gluon propagator obtained in SU(3) lattice QCD:

\begin{enumerate}
\item 
The gluon propagator $D(r)$ in the Landau gauge 
is well described 
by the four-dimensional (4D) Yukawa-type function as \cite{ISI09}
\begin{equation}
D(r) \equiv \frac{1}{24} D_{\mu\mu}^{aa}(r)= Am \frac{e^{-mr}}{r},
\end{equation}
with $m \simeq$ 600MeV and $A \simeq$ 0.16, 
in the whole region of $r \equiv (x_\alpha x_\alpha)^{1/2}=0.1\sim 1.0$fm.
\item
The gluon propagator $\tilde{D}(p^2)$ in the momentum space 
is also well described by the corresponding new-type propagator 
(4D Fourier transformed Yukawa-type function) as \cite{ISI09}
\begin{eqnarray}
\tilde{D}(p^2) =\frac{1}{24} \tilde{D}_{\mu\mu}^{aa}(p^2)
= \frac{4\pi^2Am}{(p^2+m^2)^{3/2}},
\end{eqnarray}
with $m \simeq$ 600MeV and $A \simeq$ 0.16,
in the momentum region of $0.5{\rm GeV} \le p \le 3{\rm GeV}$.
\end{enumerate}

\section{\label{sec:effmassanalytic} 
Analytical applications of Yukawa-type gluon propagator}

In this section, as applications of the Yukawa-type gluon propagator, 
we derive analytical expressions for 
the zero-spatial-momentum propagator $D_0(t)$, 
the effective mass $M_{\rm eff}(t)$, 
and the spectral function $\rho(\omega)$ of the gluon field \cite{ISI09}. 
All the derivations can be analytically performed, 
starting from the Yukawa-type gluon propagator $D_{\rm Yukawa}(r)$.

\subsection{Zero-spatial-momentum propagator of gluons}

First, we consider the zero-spatial-momentum propagator $D_0(t)$, 
associated with the Yukawa-type propagator $D_{\rm Yukawa}(r)$, 
where $r$ is the 4D Euclidean distance, $r = \sqrt{\vec{x}^2+t^2}$.
We define the zero-spatial-momentum propagator $D_0(t)$ of gluons as
\begin{eqnarray}
D_0(t) \equiv \frac{1}{24} \sum_{\vec{x}} 
\langle A_\mu^a(\vec{x},t) A_\mu^a(\vec{0},0)\rangle
= \sum_{\vec{x}} D(r),~~~
\label{eq:ZMPcorrL}
\end{eqnarray}
where the spatial momentum is projected to be zero.
For the simple argument, we here deal with the continuum formalism 
with infinite space-time. 
Starting from the Yukawa-type gluon propagator, 
\begin{equation}
D_{\rm Yukawa}(r) =\frac{Am}{r} e^{-mr}
=\frac{Am}{\sqrt{\vec{x}^2+t^2}} e^{-m\sqrt{\vec{x}^2+t^2}},
\end{equation} 
we derive the zero-spatial-momentum propagator as \cite{ISI09}
\begin{equation}
D_0(t) 
=\int d^3 x \ D_{\rm Yukawa}(r)
=4\pi Am \int_0^{\infty} dx 
\frac{x^2}{\sqrt{x^2+t^2}} e^{-m\sqrt{x^2+t^2}}
=4\pi A t K_1(mt).
\label{eq:contZMP}
\end{equation}
In Fig.2(a), we show the theoretical curve of $D_0(t)$ 
in Eq.(\ref{eq:contZMP}) with $m$=0.624GeV and $A$=0.162, 
together with the lattice QCD result of $D_0(t)$ in the Landau gauge. 
For the actual comparison with the lattice data, 
we take account of the temporal periodicity \cite{ISI09}.
The lattice QCD data are found to be well described by  
the theoretical curve, associated with the Yukawa-type gluon propagator. 

\begin{figure}[h]
\begin{center}
\includegraphics[scale=0.95]{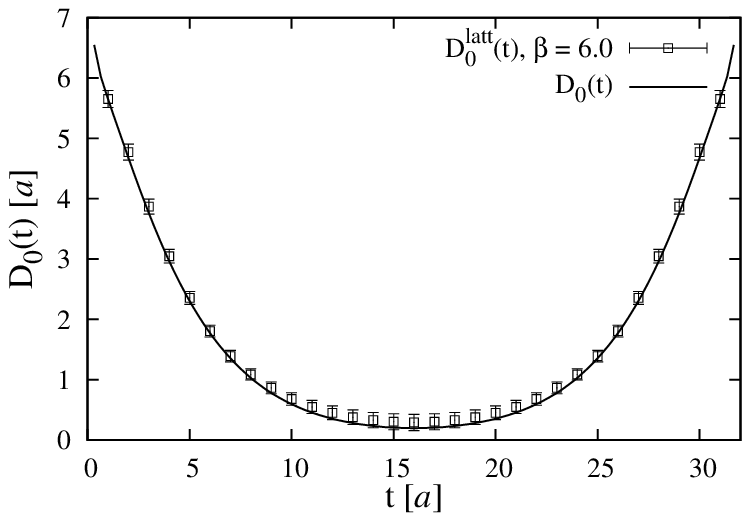}
\includegraphics[scale=0.95]{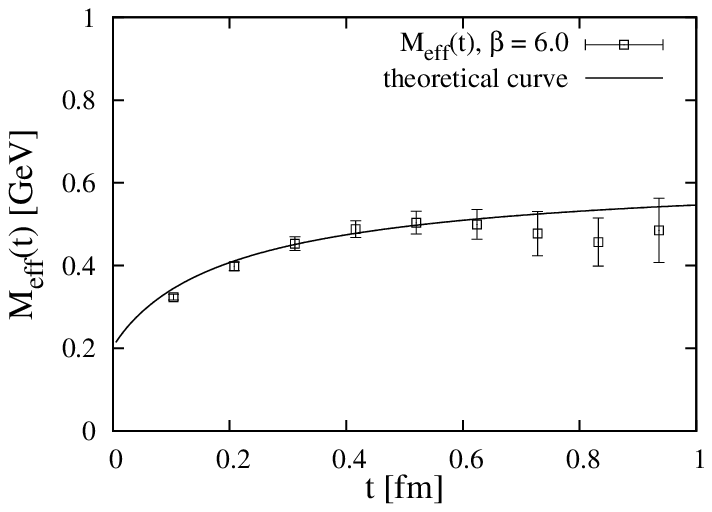}
\caption{
(a) The zero-spatial-momentum propagator $D_0(t)$ of gluons 
in the Landau gauge.
(b) The effective mass $M_{\rm eff}(t)$ of gluons in the Landau gauge.
The symbols are the lattice QCD data on $32^4$ at $\beta = 6.0$,
and the solid line is the theoretical curve derived 
from the Yukawa-type propagator with $m$=0.624GeV and $A$=0.162.
}
\end{center}
\end{figure}

\subsection{Effective mass of gluons}

Second, we investigate the effective mass $M_{\rm eff}(t)$ of gluons.
This method is often used for hadrons as a standard mass measurement 
in lattice QCD \cite{R05}. 
For the simple notation, we use the lattice unit of $a=1$ in this subsection.
In the case of large temporal lattice size, 
the effective mass of gluons is defined as 
\begin{equation}
M_{\rm eff}(t) = \ln \{D_0(t)/D_0(t+1)\}.
\label{eq:EMP}
\end{equation}

In Fig.2(b), we show the lattice result of $M_{\rm eff}(t)$, 
where we take account of the temporal periodicity.
The effective gluon mass exhibits a significant scale-dependence, 
and it takes a small value at short distances.
Quantitatively, the effective gluon mass is estimated 
to be about $400 \sim 600$MeV 
in the infrared region of about 1fm \cite{ISI09}.
This value seems consistent with the gluon mass suggested 
by Cornwall \cite{C8207}, from a systematic analysis of 
nonperturbative QCD phenomena.

Now, we consider the consequence of the Yukawa-type propagator 
$D_{\rm Yukawa}(r)$.
For simplicity, we here treat the three-dimensional space 
as a continuous infinite-volume space, 
while the temporal variable $t$ is discrete.
When the temporal periodicity can be neglected, 
we obtain an analytical expression of 
the effective mass \cite{ISI09}, 
\begin{equation}
M_{\rm eff}(t) = \ln \frac{D_0(t)}{D_0(t+1)}
=\ln \frac{t K_1(mt)}{(t+1)K_1(m(t+1))}.
\label{eq:EMGYukawa}
\end{equation}
In Fig.2(b), we add by the solid line 
the theoretical curve of $M_{\rm eff}(t)$ in 
Eq.(\ref{eq:EMGYukawa}) with $m$=0.624GeV.
The lattice QCD data of $M_{\rm eff}(t)$ are found to be well described by  
the theoretical curve derived from the Yukawa-type gluon propagator.

From the asymptotic form $K_1(z) \propto z^{-1/2}e^{-z}$, 
the effective mass of gluons is approximated as 
\begin{equation}
M_{\rm eff}(t) \simeq 
m - \frac{1}{2} \ln \big( 1 + \frac{1}{t} \big)
\simeq  m - \frac{1}{2t}
\label{eq:meffAsymptotic}
\end{equation}
for large $t$ \cite{ISI09}. 
This functional form indicates that 
$M_{\rm eff}(t)$ is an increasing function and approaches $m$ 
from below, as $t$ increases.
Then, the mass parameter $m \simeq$ 600MeV 
in the Yukawa-type gluon propagator has a definite physical meaning 
of the effective gluon mass in the infrared region.

Note that the simple analytical expression  
reproduces the anomalous increasing behavior of 
the effective mass $M_{\rm eff}(t)$ of gluons.
Thus, this framework with the Yukawa-type gluon propagator  
gives an analytical and quantitative method, and 
is found to well reproduce the lattice QCD result.

\subsection{Spectral function of gluons in the Landau gauge}

As a general argument, 
an increasing behavior of the effective mass $M_{\rm eff}(t)$ 
means that the spectral function 
is not positive-definite \cite{MO87,BPS94,MMST9395}.
More precisely, the increasing property of $M_{\rm eff}(t)$ can be realized, 
only when there is some suitable coexistence of positive- and negative-value 
regions in the spectral function $\rho(\omega)$ \cite{ISI09}.
However, the functional form of the spectral function 
of the gluon field is not yet known.

From the analytical expression of the zero-spatial-momentum propagator  
$D_0(t)=4\pi A t K_1(mt)$, 
we can derive the spectral function $\rho(\omega)$ of the gluon field, 
associated with the Yukawa-type gluon propagator \cite{ISI09}.
For simplicity,
we take continuum formalism with infinite space-time.

The relation between the spectral function $\rho(\omega)$ and 
the zero-spatial-momentum propagator $D_0(t)$ 
is given by the Laplace transformation, 
\begin{equation}
D_0(t) = \int_0^\infty d\omega \ \rho(\omega) \ e^{-\omega t}.
\label{eq:CorrvsSF}
\end{equation}
When the spectral function is given by a $\delta$-function 
such as $\rho(\omega) \sim \delta(\omega-\omega_0)$, 
which corresponds to a single mass spectrum, 
one finds a familiar relation of $D_0(t) \sim e^{-\omega_0t}$.
For the physical state, the spectral function $\rho(\omega)$ gives 
a probability factor, and is non-negative definite 
in the whole region of $\omega$.
This property is related to the unitarity of the S-matrix.

From an integral representation of the modified Bessel function, 
we derive the following formulae on the inverse Laplace transformation,
\begin{equation}
\frac{1}{2\pi i}\int_{c-i\infty}^{c+i\infty} dt \ e^{\omega t} \ K_1(t)
=\frac{\omega}{(\omega^2-1)^{1/2}} \theta(\omega-1-\varepsilon), 
\end{equation}
\vspace{-0.4cm}
\begin{eqnarray}
\frac{1}{2\pi i}\int_{c-i\infty}^{c+i\infty} dt \ e^{\omega t} \ t K_1(t)
=-\frac{1}{(\omega^2-1)^{3/2}} \theta(\omega-1-\varepsilon) 
+\frac{1}{\{2(\omega-1)\}^{1/2}} \delta(\omega-1-\varepsilon),
\end{eqnarray}
where an infinitesimal positive $\varepsilon$ is introduced 
for a regularization \cite{ISI09}.
Then, starting from the Yukawa-type propagator,
we derive the spectral function $\rho(\omega)$ 
of the gluon field as \cite{ISI09} 
\begin{equation}
\rho(\omega)
= -\frac{4\pi A m}{(\omega^2-m^2)^{3/2}}\theta(\omega-m-\varepsilon)
+\frac{4\pi A/\sqrt{2m}}{(\omega-m)^{1/2}} \delta(\omega-m-\varepsilon),
\label{eq:SFYukawa}
\end{equation}
with an infinitesimal positive $\varepsilon$. 
Here, $m \simeq$ 600MeV is the mass parameter in the Yukawa-type function  
for the Landau-gauge gluon propagator.
The first term expresses a negative continuum spectrum, and 
the second term a $\delta$-functional peak with 
the residue including a positive infinite factor as $\varepsilon^{-1/2}$ 
at $\omega=m+\varepsilon$.

\begin{figure}[h]
\begin{center}
\includegraphics[scale=1.1]{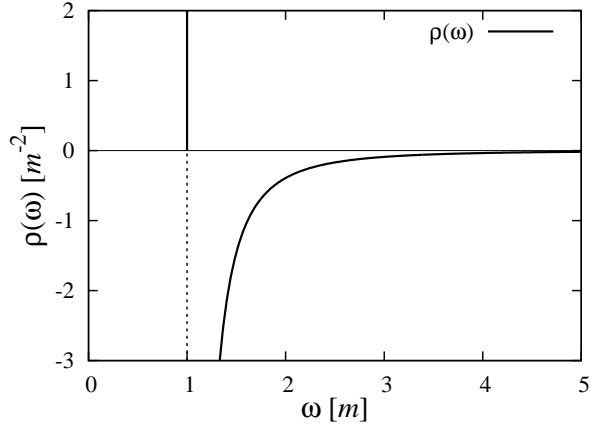}
\caption{\label{fig:Spectral}
The spectral function $\rho(\omega)$ of the gluon field, 
associated with the Yukawa-type propagator.
The unit is normalized by the mass parameter $m \simeq$ 600MeV. 
$\rho(\omega)$ shows anomalous behaviors: 
it has a positive $\delta$-functional peak 
with the residue of $+\infty$ at $\omega = m$, 
and takes negative values for all the region of $\omega > m$.
}
\end{center}
\vspace{-0.1cm}
\end{figure}

We show in Fig.3 the spectral function $\rho(\omega)$ of the gluon field.
As a remarkable fact, the obtained gluon spectral function $\rho(\omega)$ 
is negative-definite for all the region of $\omega > m$, 
except for the positive $\delta$-functional peak at $\omega=m$.
The negative property of the spectral function in coexistence 
with the positive peak leads to the anomalous increasing behavior 
of the effective mass $M_{\rm eff}(t)$ of gluons \cite{ISI09}.
Actually, the resulting effective mass $M_{\rm eff}(t)$ 
well describes the lattice result, as shown in Fig.2(b).

We note that the gluon spectral function $\rho(\omega)$ is divergent 
at $\omega=m+\varepsilon$, 
and the divergence structure consists of two ingredients: 
a $\delta$-functional peak with a positive infinite residue 
and a negative wider power-damping peak.
On finite-volume lattices, these singularities are to be smeared, 
and $\rho(\omega)$ is expected to take a finite value everywhere on $\omega$. 
On the lattice, we conjecture that the spectral function $\rho(\omega)$ 
includes a narrow positive peak stemming from the $\delta$-function 
in the vicinity of $\omega=m \ (+\varepsilon)$  
and a wider negative peak near $\omega \simeq m$ 
in the region of $\omega > m$ \cite{ISI09}.

In this way, the Yukawa-type gluon propagator indicates 
an extremely anomalous spectral function 
of the gluon field in the Landau gauge. 
The obtained gluon spectral function $\rho(\omega)$ 
is negative almost everywhere, and 
includes a complicated divergence structure 
near the ``anomalous threshold'', $\omega=m\ (+\varepsilon)$. 
Thus, this framework with the Yukawa-type gluon propagator 
gives an analytical and concrete expression 
for the gluon spectral function $\rho(\omega)$ at the quantitative level.

\section{A hypothesis of an effective dimensional reduction 
in stochastic gluonic vacuum by the Parisi-Sourlas mechanism}

We discuss the Yukawa-type gluon propagation and 
a possible dimensional reduction 
due to the stochastic behavior of the gluon field 
in the infrared region \cite{ISI09}. 
As shown before, the Landau-gauge gluon propagator is well described 
by the Yukawa function in {\it four}-dimensional Euclidean space-time. 
However, the Yukawa function $e^{-mr}/r$ is a natural form 
in {\it three}-dimensional Euclidean space-time, 
since it is obtained by the three-dimensional Fourier transformation of 
the ordinary massive propagator $(p^2+m^2)^{-1}$. 
In fact, the Yukawa-type propagator has a ``three-dimensional'' property.
In this sense, as an interesting possibility, 
we propose to interpret this Yukawa-type behavior of the gluon propagation 
as an ``effective reduction of the space-time dimension''.

Such a ``dimensional reduction'' sometimes occurs in stochastic systems, 
as Parisi and Sourlas pointed out 
for the spin system in a random magnetic field \cite{PS79}.
In fact, on the infrared dominant diagrams, the $D$-dimensional system 
coupled to the Gaussian-random external field 
is equivalent to the $(D-2)$-dimensional system without the external field.  

\begin{figure}[h]
\vspace{0.4cm}
\begin{center}
\includegraphics[scale=0.83]{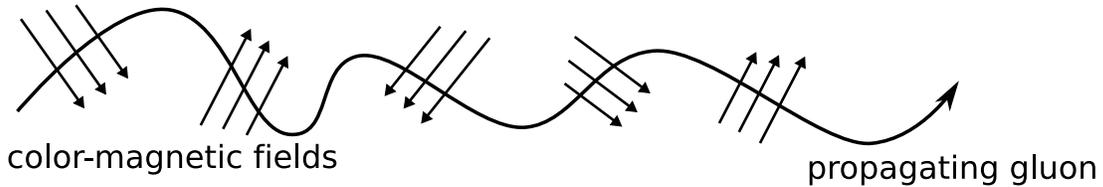}
\caption{
A schematic figure for a propagating gluon in the
QCD vacuum. The QCD vacuum is filled with color-magnetic
fields which are stochastic at an infrared scale, 
and the gluon propagates in the random color-magnetic fields.
}
\end{center}
\end{figure}

We note that the gluon propagation in the QCD vacuum resembles 
the situation of the system coupled to the stochastic external field.
Actually, as is indicated by a large positive value of the gluon condensate 
$\langle G_{\mu\nu}^aG_{\mu\nu}^a\rangle >0$ in the Minkowski space,  
the QCD vacuum is filled with a strong color-magnetic field \cite{R05,S77}, 
which can contribute spontaneous chiral-symmetry breaking \cite{ST9193}, 
and the color-magnetic field is considered to be highly random 
at the infrared scale \cite{S77,IS9900}.
Since gluons interact each other, 
the propagating gluon is violently scattered 
by the other gluon fields which are randomly condensed 
in the QCD vacuum at the infrared scale, as shown in Fig.4.

Actually at the infrared scale, 
the gluon field shows a strong randomness due to the strong interaction, 
and this infrared strong randomness 
is considered to be responsible for color confinement, 
as is indicated in strong-coupling lattice QCD \cite{R05}. 
Even after the removal of the fake gauge degrees of freedom by gauge fixing, 
the gluon field exhibits a strong randomness \cite{IS9900} 
accompanying a quite large fluctuation at the infrared scale.

As a generalization of the Parisi-Sourlas mechanism, 
we conjecture that the infrared structure of a theory 
in the presence of the quasi-random external field 
in higher-dimensional space-time has a similarity to 
the theory without the external field 
in lower-dimensional space-time \cite{ISI09}.
From this point of view, the Yukawa-type behavior of gluon propagation 
may indicate an ``effective reduction of space-time dimension'' by one, 
due to the stochastic interaction between the propagating gluon 
and the other gluon fields in the QCD vacuum, 
of which net physical fluctuation is highly random at the infrared scale.

\section{Relevant gluonic scale for color confinement}

In this section, as another subject, we study lattice-QCD analysis for 
the relevant gluonic momentum-component for color confinement \cite{YS0809}. 
Here, we formulate a new general lattice framework 
to extract the relevant gluonic energy scale of each QCD phenomenon 
by introducing a cut for link-variables in momentum space \cite{YS0809}.
Our method consists of the following five steps.

\vspace{0.4cm}

\noindent{\it 
Step 1. Generation of coordinate-space link-variable in the Landau gauge
}

\vspace{0.3cm}

As usual, we generate a gauge configuration on a $L^4$ lattice with 
the lattice spacing $a$ 
by the lattice-QCD Monte Carlo simulation 
under space-time periodic boundary conditions, 
and obtain a finite number of coordinate-space link-variables.
Here, we consider the link-variables fixed in the Landau gauge, 
which gives a transparent connection between the link-variable 
and the gauge field, owing to the suppression of gluon-field fluctuations.

\vspace{0.4cm}

\noindent{\it
Step 2. Four-dimensional discrete Fourier transformation
}

\vspace{0.3cm}

By the discrete Fourier transformation, 
we define the momentum-space link-variable, 
\begin{eqnarray}
{\tilde U}_{\mu}(p)=\frac{1}{N_{\rm site}}\sum_x 
U_{\mu}(x)\exp(i {\textstyle \sum_\nu} p_\nu x_\nu),
\end{eqnarray}
where $N_{\rm site}$ is the total number of lattice sites.
The momentum-space lattice spacing is given by
\begin{eqnarray}
a_p = \frac{2\pi}{La}.
\end{eqnarray}

\vspace{0.4cm}

\noindent{\it
Step 3. Imposing a cut in the momentum space
}

\vspace{0.3cm}

We impose a cut on ${\tilde U}_{\mu}(p)$ in a certain region 
of the momentum space, as schematically shown in Fig.5. 
Outside the cut, we replace $\tilde U_{\mu}(p)$ 
by the free-field link-variable,
${\tilde U}^{\rm free}_{\mu}(p)=
\frac{1}{N_{\rm site}}\sum_x 1 
\exp(i {\textstyle \sum_\nu} p_\nu x_\nu)=\delta_{p0}$, 
corresponding to $A_\mu(x)=0$ or $U_\mu(x)=1$.
Then, the momentum-space link-variable 
${\tilde U}_{\mu}^{\Lambda}(p)$ 
with the cut is defined as 
\begin{equation}
\label{eq2}
{\tilde U}_{\mu}^{\Lambda}(p)= \Bigg\{
\begin{array}{cc}
{\tilde U}_{\mu}(p) & ({\rm inside \ cut})\\
{\tilde U}^{\rm free}_{\mu}(p)=\delta_{p0} & ~~({\rm outside \ cut}).
\end{array}
\end{equation}

\begin{figure}[h]
\vspace{-0.1cm}
\begin{center}
\includegraphics[scale=0.38]{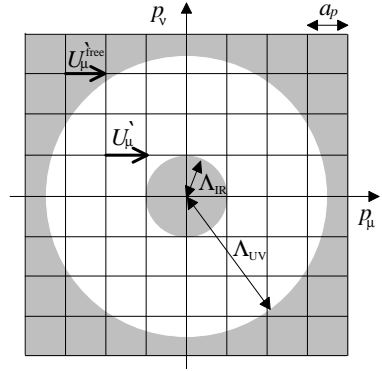}
\caption{
A schematic figure of the UV cut $\Lambda_{\rm UV}$ 
and the IR cut $\Lambda_{\rm IR}$ on momentum-space lattice, 
with the lattice spacing $a_p=2\pi/(La)$.
The momentum-space link-variable ${\tilde U}_{\mu}(p)$ is replaced 
by the free variable ${\tilde U}^{\rm free}_{\mu}(p)$ 
in the shaded cut regions.
}
\end{center}
\vspace{-0.2cm}
\end{figure}

\vspace{0.4cm}

\noindent{\it
Step 4. Inverse Fourier transformation
}

\vspace{0.3cm}

To return to coordinate space, 
we carry out the inverse Fourier transformation as
\begin{eqnarray}
U'_{\mu}(x)=\sum_p {\tilde U}_{\mu}^{\Lambda}(p)
\exp(-i {\textstyle \sum_\nu} p_\nu x_\nu).
\end{eqnarray}
Since this $U'_{\mu}(x)$ is not an SU(3) matrix, 
we project it onto an SU(3) element $U^{\Lambda}_{\mu}(x)$ by maximizing
$
{\rm ReTr}[U^{\Lambda}_{\mu}(x)^{\dagger}U'_{\mu}(x)].
$
Such a projection is often used in lattice QCD algorithms.
By this projection, we obtain the coordinate-space link-variable 
$U^{\Lambda}_{\mu}(x)$ with the cut, 
which is an SU(3) matrix and has the maximal overlap to $U'_{\mu}(x)$.

\vspace{0.4cm}

\noindent{\it
Step 5. Calculation of physical quantities
}

\vspace{0.3cm}

Using the cut link-variable $U^{\Lambda}_{\mu}(x)$, instead of $U_{\mu}(x)$, 
we calculate physical quantities as the expectation value
in the same way as original lattice QCD.

\vspace{0.55cm}

With this method in lattice-QCD framework, we quantitatively determine 
the relevant energy scale of color confinement, 
through the analyses of the $Q\bar Q$ potential.
The lattice QCD Monte Carlo simulations are performed on $16^4$ lattice 
at $\beta$=5.7, 5.8, and 6.0 at the quenched level \cite{YS0809}.

\begin{figure}[h]
\vspace{-0.35cm}
\begin{center}
\includegraphics[scale=0.85]{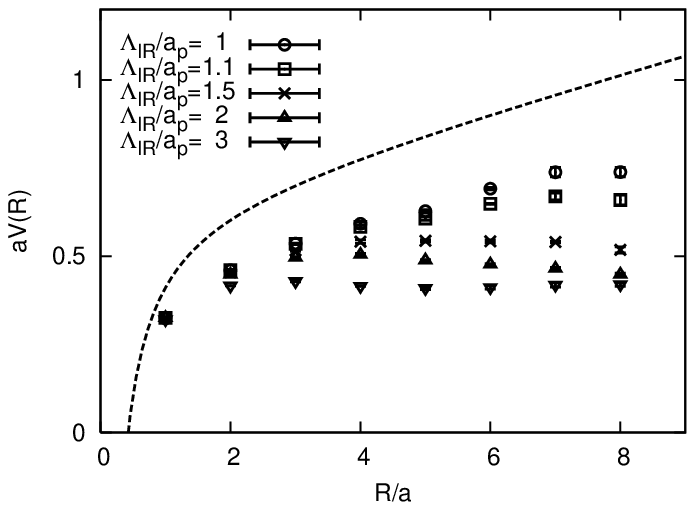}
\includegraphics[scale=0.85]{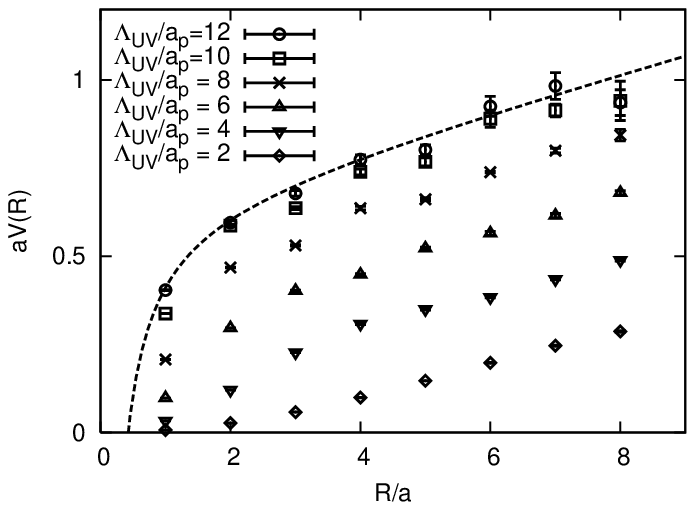}
\vspace{-0.15cm}
\caption{
(a) The $Q\bar Q$ potential $V(R)$ with the IR cut 
$\Lambda_{\rm IR}$ plotted against the interquark distance $R$.
(b) The $Q\bar Q$ potential with the UV cut $\Lambda_{\rm UV}$.
The lattice QCD calculation is performed on $16^4$ lattice with $\beta =6.0$, 
{\it i.e.}, $a\simeq 0.10$ fm and $a_p \equiv 2\pi/(La) \simeq 0.77$ GeV.
The broken line is the original $Q\bar Q$ potential in lattice QCD.
}
\end{center}
\vspace{-0.4cm}
\end{figure}

Figure 6 (a) and (b) show the $Q\bar Q$ potential $V(R)$ 
with the IR cutoff $\Lambda_{\rm IR}$ 
and the UV cutoff $\Lambda_{\rm UV}$, respectively.
We get the following lattice-QCD results 
on the role of gluon momentum components.
\begin{itemize}
\item
By the IR cutoff $\Lambda_{\rm IR}$, as shown in Fig.6(a), 
the Coulomb potential seems to be unchanged, 
but the confinement potential is largely reduced \cite{YS0809}.
\item
By the UV cutoff $\Lambda_{\rm UV}$, as shown in Fig.6(b), 
the Coulomb potential is largely reduced, 
but the confinement potential is almost unchanged \cite{YS0809}.
\end{itemize}

\begin{figure}[h]
\vspace{-0.1cm}
\begin{center}
\includegraphics[scale=0.9]{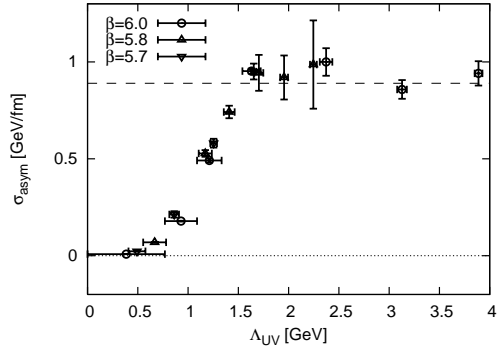}
\vspace{-0.1cm}
\caption{
The $\Lambda_{\rm UV}$-dependence of the string tension 
$\sigma$ obtained from 
the asymptotic slope of the $Q\bar Q$ potential 
$V(R)$ with the UV cutoff $\Lambda_{\rm UV}$. 
The lattice QCD calculations are performed on $16^4$ lattice 
with $\beta$ =5.7, 5.8 and 6.0. 
The vertical error-bar is the statistical error, 
and the horizontal error-bar the range from 
the discrete momentum.
The broken line denotes the original value of 
the string tension $\sigma \simeq 0.89$ GeV/fm.
}
\end{center}
\vspace{-0.2cm}
\end{figure}

Fig.7 shows the $\Lambda_{\rm UV}$-dependence of the string tension 
$\sigma$ obtained from the asymptotic slope of the $Q\bar Q$ potential 
$V(R)$ with the UV cutoff $\Lambda_{\rm UV}$.
As a remarkable fact, the string tension is almost unchanged 
even after cutting off the high-momentum gluon component above 1.5 GeV.
In fact, the relevant gluonic scale of color confinement is concluded to be 
below 1.5 GeV \cite{YS0809}. 

\section*{Acknowledgements}
\vspace{-0.1cm}
H.S. is grateful to Prof. J.M.~Cornwall for useful suggestions. 
He thanks the organigers of QCD-TNT.
This work is supported by a Grant-in-Aid for Scientific Research 
[(C) No.~19540287] in Japan.
The lattice QCD calculations are done on NEC SX-8R at Osaka University.

\end{document}